\documentclass[fleqn,10pt]{SelfArx} 

\definecolor{color1}{RGB}{0,0,90} 
\definecolor{color2}{RGB}{0,20,20} 

\usepackage{hyperref} 
\hypersetup{hidelinks,colorlinks,breaklinks=true,urlcolor=color2,citecolor=color1,linkcolor=color1,bookmarksopen=false,pdftitle={Title},pdfauthor={Author},urlcolor=blue}

\usepackage{graphicx}
\usepackage[square]{natbib}
\usepackage{amsmath}
\usepackage{multirow}
\usepackage{subfigure}

\newcommand{\n}[1]{\mathrm{#1}}

\JournalInfo{Published in Applied Physics Letters, Vol. 103 (10), 102403, 2013} 
\Archive{\href{http://dx.doi.org/10.1063/1.4820141}{DOI: 10.1063/1.4820141}} 

\PaperTitle{Demagnetization factor for a powder of randomly packed spherical particles} 

\Authors{R. Bj\o{}rk and C.R.H. Bahl} 
\affiliation{\textit{Department of Energy Conversion and Storage, Technical University of Denmark - DTU, Frederiksborgvej 399, DK-4000 Roskilde, Denmark}} 
\affiliation{*\textbf{Corresponding author}: rabj@dtu.dk} 

\newcommand{\keywordname}{Keywords} 

\Abstract{The demagnetization factors for randomly packed spherical particle powders with different porosities, sample aspect ratios and monodisperse, normal and log-normal particle size distributions have been calculated using a numerical model. For a relative permeability of 2, comparable to room temperature Gd, the calculated demagnetization factor is close to the theoretical value. The normalized standard deviation of the magnetization in the powder was 6.0\%-6.7\%. The demagnetization factor decreased significantly, while the standard deviation of the magnetization increased, for increasing relative permeability.}

\begin{document}

\flushbottom 

\maketitle 


\thispagestyle{empty} 

Measuring the magnetic response of a sample when it is subjected to a uniform external magnetic field is fundamental in studying the magnetic properties of a material. For ellipsoidal bodies the internal magnetic field in the material will be uniform when subjected to a uniform external field, and can be calculated analytically \cite{Osborn_1945}. This is also the case for an infinite sheet or an infinite cylinder, but in general this is not possible for other geometrical shapes.

In general the internal magnetic field, $\mathbf{H}$, in a sample can be expressed in terms of the externally applied field $\mathbf{H}_\textrm{appl}$, as
\begin{eqnarray}\label{Eq.Demag_def}
\mathbf{H}=\mathbf{H}_\textrm{appl}-\mathbb{N}\cdot{}\mathbf{M}
\end{eqnarray}
where $\mathbf{M}$ is the magnetization and $\mathbb{N}$ is the demagnetization tensor. Usually, this equation is expressed as a scalar equation where the demagnetization tensor is replaced by an average demagnetization factor, $N$.
For a cuboid sample\footnote{Also known as rectangular parallelepiped or simply rectangular box.}, the magnetization throughout the sample will not be uniform, but an approximate analytical expression exists for the average demagnetization factor \cite{Aharoni_1998}, which has also been verified numerically \cite{Smith_2010}. The non-homogeneous demagnetization field can be  visualized directly using thermography and the magnetocaloric effect\cite{Christensen_2010}. For packed sphere powders, the average demagnetization factor has been calculated for regular packings, i.e. simple-cubic, body-centered-cubic and face-centered-cubic packed spheres \cite{How_1991}. Assuming homogeneous magnetization, it has also been calculated for aggregates of particles in a matrix \cite{Skomski_2007}. Besides these geometries, the demagnetization factors for various two-dimensional arrays of magnetic materials \cite{Martinez_Huerta_2013} as well as for cylinders \cite{Chen_2006} have also been calculated numerically, and the influence of demagnetization for nano-sized materials has also been investigated \cite{Skomski_2010}.

For a powder of randomly packed spherical particles, Breit \cite{Breit_1922} and Bleany \cite{Bleaney_1941} have both shown that the demagnetization factor may be expressed as
\begin{eqnarray}\label{Eq.Demag}
N=\frac{1}{3}+f\left(D_z-\frac{1}{3}\right)
\end{eqnarray}
where $D_z$ is the demagnetization factor of the geometrical shape of the powder sample itself and $f$ is the relative density or packing fraction. It is assumed that the spheres are all uniformly magnetized. In the limit of full density, Eq. (\ref{Eq.Demag}) reduces to the demagnetization factor of the geometry of the sample. In the case of very low packing density, where the spheres do not interact magnetically, the demagnetization factor reduces to 1/3, which is that of an isolated sphere. There is no experimental or numerical verification of Eq. (\ref{Eq.Demag}) for intermediate packing fractions. Also, it is not known if Eq. (\ref{Eq.Demag}) is correct for e.g. powders with different particle size distributions or particle shapes, nor if the spheres are not uniformly magnetized.

Knowing the demagnetization factor for randomly packed spheres is important when characterizing magnetic materials, where the samples often come in the form of loose powders. Containers of tightly packed spheres are also frequently used in e.g. magnetic refrigeration devices, where they are subjected to an alternating magnetic field \cite{Yu_2010}. To accurately model this process the average demagnetization factor of the packed spheres must be known.

Here we present numerical modeling results of the magnetization of packed spheres with different particle size distributions and sample aspect ratios. The objective of this study is to compare Eq. (\ref{Eq.Demag}) to the actual demagnetization factor calculated, when the simplifying assumption of constant magnetization is dropped, i.e. as would be the case in real world materials.

We consider three different particle size distributions for the packed spheres; monodisperse, normal and log-normal.  The normal and log-normal particle size distributions are included as powder and particle preparation techniques can result in both distributions \cite{Patterson_1985}. The size of the monodisperse particles are equal to the average particle size of the normal distribution and the mode of the log normal distribution. The latter two distributions have a standard deviation and skewness of one, respectively.

For each of the different distributions three different aspect ratios of the sample container are considered; 1, 1.5 and 2. For each set of particle size distribution and aspect ratio, seven different relative densities of the powder samples are considered. We consider relative densities in the range 0.4-0.6, as this is the corresponding range for very loose to tightly packed powders. A sphere packing generated by slow settling in a fluidized bed produces a density of 0.56; dropping spheres into a container results in a density of 0.6-0.61 \cite{Dullien_1992}. The lower densities considered are for magnetic powders suspended in a non-magnetic matrix, e.g. in an organic slurry used for tapecasting magnetic structures \cite{Bahl_2012}.

All length scales are given in arb. units, as magnetostatics problems are scale invariant, i.e. if all dimensions are scaled by the same factor the magnetic field in a given position will be the same if this position is scaled as well. This means that quantities such as the average internal field in a scaled volume of space will be the same, and therefore the length unit is arbitrary. We only assume that domains and domain walls are much smaller than the macroscopic size of the spheres, thus the results are not applicable to nano-spheres \cite{Skomski_2010}. The results will thus be applicable to most systems of packed spheres, albeit one usually considers spheres in the sub-millimeter to millimeter range, packed into a powder. While the magnetostatics problems are invariant in scale for the same systems, the internal scales in a sample might have an influence on the magnetic response of the sample. Therefore, it is important to investigate the response of samples with different particle size distributions.

The powder samples used for calculating the demagnetization factor have been generated by simulating the pouring of spherical particles of different particle size distributions into a cubic container consisting of $200\times{}200\times{}n_z$ voxels, where $n_z$ is equal to 200, 133 and 100 for the three different aspect ratios considered here. The aspect ratio is thus defined as $n_x/n_z$. The number of particles in a given powder sample varies between 200 and 1000, with the number being lowest for the log-normal distributions with an aspect ratio of 2.

\begin{figure}[!tp]
  \centering
  \includegraphics[width=0.7\columnwidth]{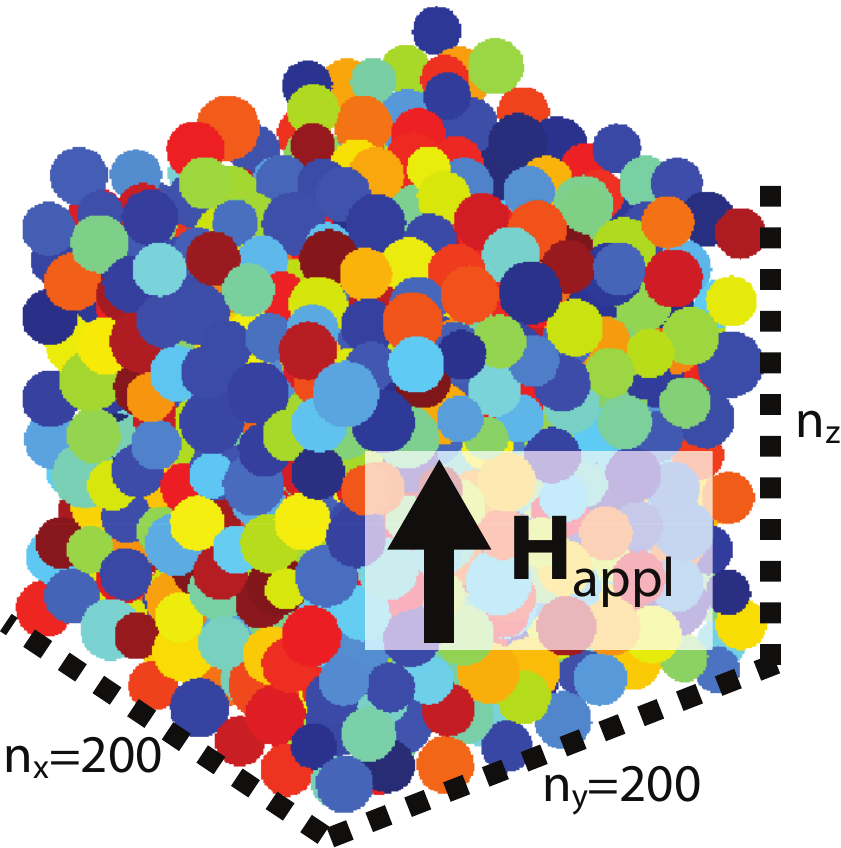}
  \caption{[Color online] The powder of packed spherical particles with a normal distribution and with a relative density of the powder sample of 0.52. All spheres have been given a unique color. The aspect ratio of the sample is 1. The dimensions of the sample have been indicated, as well as the direction of the applied field, which is along $n_z$.}
    \label{Fig.Demag_particles}
\end{figure}

The numerical code used to simulate the pouring and packing of the spherical particles is a modified version of the Large-scale Atomic/Molecular Massively Parallel Simulator (LAMMPS) code, available as open source from Sandia National Laboratories. A random seed is used in the generation of the poured spheres, meaning different powder samples can be generated for the same distribution parameters. In order to simulate powders with different porosities, small ''pore former`` particles with a radius of half the size of the monodisperse particles were added to the poured spheres. After the simulated pouring, these particles are removed, leaving behind pores and thereby increased porosity. An illustration of the powder sample with an aspect ratio of one and a normal distribution of spheres with a density of 0.52 is shown in Fig. \ref{Fig.Demag_particles}.


For each set of particle size distribution, aspect ratio and porosity, twenty different packed sphere powders have been generated in order to calculate the statistical uncertainty on the computed demagnetization factor. The demagnetization factor has been computed using the commercially available finite element software tool Comsol Multiphysics \citep{comsol}. The system modeled is a static problem in magnetism, i.e. magnetostatics. The equation solved is the magnetic scalar potential equation, which is a Laplace's equation,
\begin{eqnarray}
-\nabla{}\cdot{}(\mu{}_{0}\mu{}_{r}\nabla{}V_\mathrm{m})=0~,\label{Eq.Numerical_Magnetism}
\end{eqnarray}
where $\mu{}_{0}$ is the permeability of free space, $\mu{}_{r}$ is the relative permeability, which in all cases is assumed to be constant and isotropic, and $V_\mathrm{m}$ is the magnetic scalar potential from which the magnetic field can be found as $-\nabla{}V_\mathrm{m} + \mathbf{H}_\textrm{appl} = \mathbf{H}$, where $\mathbf{H}_\textrm{appl}$ is the applied magnetic field, assumed here to be $\mu_0\mathbf{H}_\textrm{appl}=1$ T in the direction along the height of the sample container, as also shown in Fig. \ref{Fig.Demag_particles}. The above equation is solved on a finite element mesh and the solver used is \emph{Pardiso} which is a parallel sparse direct linear solver \citep{Schenk_2001,Schenk_2002}. The computational volume is chosen large enough that the boundaries do not affect the calculations.

The spherical particles are modelled as linear materials with a constant $\mu_r$. We have found numerically that for such materials the demagnetization factor, $N$, is independent of the applied magnetic field, and thus independent of the actual value of the magnetisation at a given applied field. This means that the demagnetization factor will only depend on the response of the magnetization to the applied field, i.e. to $\chi_\textrm{d} =\partial{}M/\partial{}H_\textrm{appl}$. As the applied field is increased $\chi_\textrm{d}$ for a ferromagnetic material will decrease towards the saturation value of zero. Approaching saturation the value of $\chi_\textrm{d}$ changes slowly, making it a reasonable approximation to model the materials as being linear. In the results presented below, a constant relative permeability of $\mu_r = 2$ has been chosen. This is close to the permeability of gadolinium at room temperature in a 1 T field \cite{Bjoerk_2010d}, making the calculated results applicable to e.g. magnetized beds of packed spheres of gadolinium. Indeed, at an applied field of 1 T the permeability of most ferromagnetic materials is of this order. The effect of varying the value of $\mu_r$ will be considered subsequently.

Estimating the relative density or packing fraction of the generated powder samples is not trivial. While the relative density will be homogeneous throughout most of the sample, it will decrease near the edges due to the surface effect. Here the relative density has been taken to be the relative density in the central part of the sample, taken here to be the inner 80\% of the volume of the sample. Thus the effect of the surface on the density of the sample is ignored.

\begin{figure}[!tp]
  \centering
  \includegraphics[width=1\columnwidth]{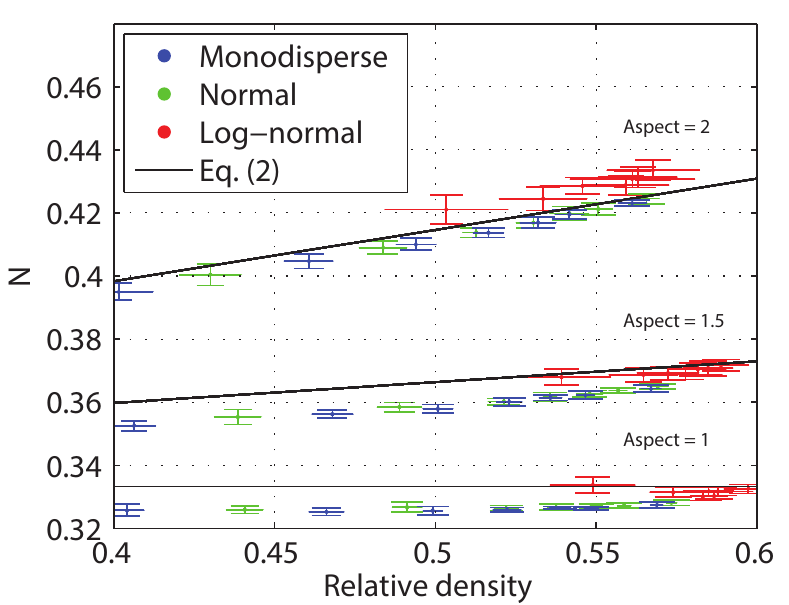}
  \caption{[Color online] The demagnetization factor as function of relative density for the monodisperse, normal and log-normal distributions considered, for different aspect ratios. The error bars are the standard deviation calculated for the twenty different samples with the same particle size distribution, aspect ratio and porosity}
    \label{Fig.Demag}
\end{figure}

The calculated average demagnetization factor, $N$, is shown in Fig. \ref{Fig.Demag} for the three different particle size distributions as function of relative density. The error bars indicate the standard deviation of both the porosity and the calculated demagnetization factor for the twenty different samples with the same particle size distribution, aspect ratio and porosity. As can clearly be seen from the figure, the calculated demagnetization factors very closely follow the trend of Eq. (\ref{Eq.Demag}), where the $D_z$ factor is calculated using the approximate analytical expression for the demagnetization of a cuboid \cite{Aharoni_1998}. Also, the demagnetization factor does not appear to be a strong function of the particle size distribution, at least for the distribution widths considered here. This is important experimentally, as real powder samples almost always have a particle size distribution of some width. The slight deviation from Eq. (\ref{Eq.Demag}) could be caused by the assumption in the derivation of Eq. (\ref{Eq.Demag}) that the spheres are all uniformly magnetized, which is not the case in the modeled sample.

\begin{figure}[!tp]
  \centering
  \includegraphics[width=1\columnwidth]{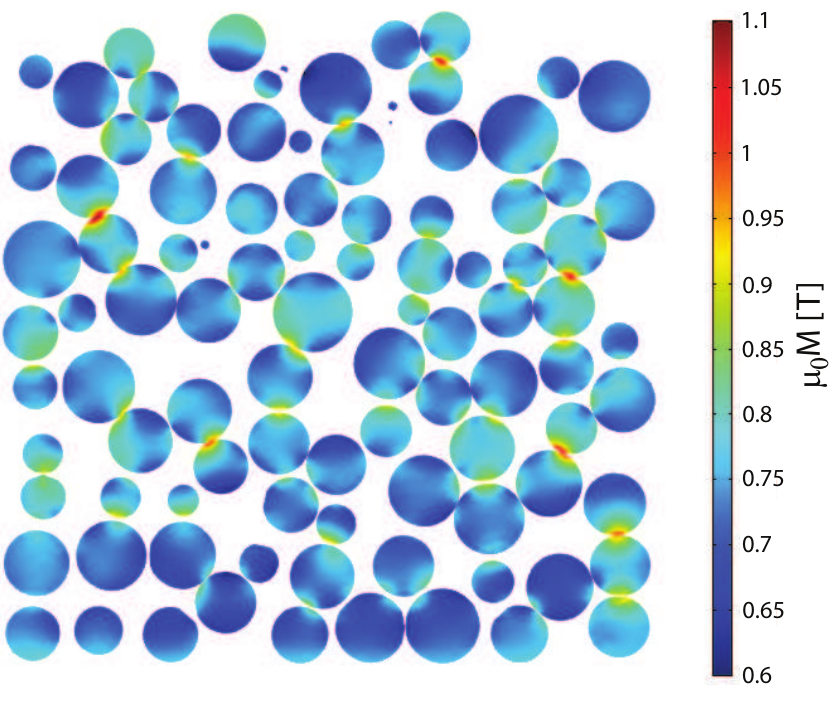}
  \caption{[Color online] A slice through the center of the sample with a normal particle size distribution with a relative density of 0.55. The color indicates the magnetization times the permeability of free space, $\mu_{0}M$.}
    \label{Fig.Magnetization_slice}
\end{figure}

It is important to evaluate whether the deviation from Eq. (\ref{Eq.Demag}) can be caused by the magnetization inside the individual spherical particles being non-homogenous. This is also important as some physical properties of the particles, such as e.g. the magnetocaloric effect, depends on the internal field in the spheres. The normalized standard deviation, $\sigma$, of the magnetization, $M$, has been calculated inside every single sphere and will be termed $\eta_\n{individual}$. The overall normalized standard deviation of the magnetisation within each of the powder samples is likewise termed $\eta_\n{powder}$. The standard deviation of the magnetization for the individual particles is around $\eta_\n{individual} = (5\pm 1.4)\%$ for all powders, with the spread being largest for the log-normal distributions. For the powder sample as a whole the standard deviation of the magnetization is around $\eta_\n{powder} = 6.0\%-6.7\%$. Thus the magnetization can be considered to be fairly uniform in the powder for $\mu_r = 2$. However, the standard deviation could be responsible for the deviation from Eq. (\ref{Eq.Demag}) that was seen in Fig. \ref{Fig.Demag}. The uniformity of the magnetization can be seen for one of the powder samples in Fig. \ref{Fig.Magnetization_slice}.

\begin{figure}[t]
  \centering
  \includegraphics[width=1\columnwidth]{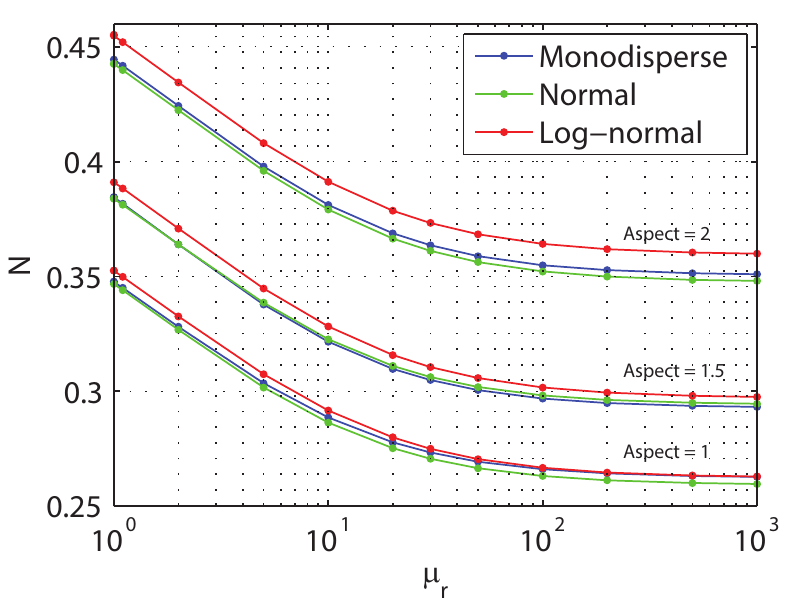}
  \caption{[Color online] The demagnetization factor as function of the relative permeability, $\mu_r$, for the highest density monodisperse, normal and log-normal distributions considered, for different aspect ratios.}
    \label{Fig.Mur_function_all}
\end{figure}

Both the analytical expression for the demagnetization factor of a cuboid, as well as the expression for a powder of packed spherical particles, assumes the magnetization to be completely parallel to the applied field. While this is a good approximation for low values of the relative permeability, the calculated demagnetization factor can be substantially different for high values of the relative permeability. In ferromagnetic materials below the Curie temperature the permeability is initially high at low applied magnetic fields, decreasing as the material gets closer to saturation. Shown in Fig. \ref{Fig.Mur_function_all} is the  demagnetization factor as function of the relative permeability. The demagnetization factor is calculated using the scalar and averaged version of Eq. (\ref{Eq.Demag_def}) for a single of the twenty samples with the highest density considered above, for all distributions and aspect ratios. As can clearly be seen, the demagnetization factor decreases significantly the higher the value of the relative permeability, until a constant value is reached for $\mu_r > 100$. This is also observed for demagnetization of cylinders \cite{Chen_2006}. Thus the correction for demagnetization in order to find the average magnetization is not purely given by Eq. (\ref{Eq.Demag}), and the deviation observed in Fig. \ref{Fig.Demag} could in part be caused by the chosen value of $\mu_r=2$, which will reduce $N$ slightly. A significant non-uniformity of the magnetization inside the sample is also present when $\mu_r$ increases. For all samples the normalized standard deviation increases to $\approx 35$\% at $\mu_r=1000$, following a curve similar to the curves in Fig. \ref{Fig.Mur_function_all}, except increasing.

In conclusion, we have calculated the demagnetization factor for a powder sample of packed spherical particles with monodisperse, normal and log-normal particle size distributions for different porosities and sample aspect ratios. The demagnetization factor is not a function of applied field, but only the change in magnetization with respect to field, i.e. $\chi_\textrm{d}$. For a relative permeability of 2, which is comparable to that of Gd at room temperature at an applied field of 1 T, the calculated demagnetization factor is close to the theoretically predicted value. The demagnetization factor does not appear to be a function of particle size distribution. The magnetizations both inside the individual spheres, as well as for the whole sample, have normalized standard deviations of 5 and 6.0\%-6.7\%, respectively. Relative permeabilities larger than 2 were also analyzed, finding that the demagnetization factor decreases significantly for high values of the permeability, while the standard deviation of the magnetization increases.


\begin{thebibliography}{10}

\bibitem{Osborn_1945}
J.~A. Osborn, Phys. Rev. {\bf 67}, 351-357 (1945).

\bibitem{Aharoni_1998}
A.~Aharoni, J. Appl. Phys. {\bf 83}(6), 3432-3434 (1998).

\bibitem{Smith_2010}
A.~Smith, K.~K. Nielsen, D.~V. Christensen, C.~R.~H. Bahl, R.~Bj\o{}rk, and J.~Hattel, J. Appl. Phys. {\bf 107}, 103910 (2010).

\bibitem{Christensen_2010}
D.V. Christensen, R.~Bj\o{}rk, K.K. Nielsen, C.R.H. Bahl, A.~Smith, D.V. Christensen, K.K. Nielsen, and S.~Clausen, J. Appl. Phys. {\bf 108}(6), 063913 (2010).

\bibitem{How_1991}
H.~How and C.~Vittoria, Phys. Rev. B {\bf 43}(10), 8094-8104 (1991).

\bibitem{Skomski_2007}
R. Skomski, G.C. Hadjipanayis, and D.J. Sellmyer, IEEE Trans. Magn. {\bf 43}(6), 2956-2958 (2007).

\bibitem{Martinez_Huerta_2013}
J.~M. Martinez-Huerta, A.~Encinas, J.~De La~Torre Medina, and L.~Piraux, J. Phys. Cond. Mat. {\bf 25}, 226003 (2013).

\bibitem{Chen_2006}
D.-X. Chen, E.~Pardo, and A.~Sanchez, J. Magn. Magn. Mater. {\bf 306}(1), 135-146 (2006).

\bibitem{Skomski_2010}
R. Skomski, Y. Liu, J. E. Shield, G. C. Hadjipanayis, and D.J. Sellmyer, J. Appl. Phys. {\bf 107}, 09A739 (2010).

\bibitem{Breit_1922}
G.~Breit. Commun. Phys. Lab. Univ. Leiden Suppl. no. 46. (1922).

\bibitem{Bleaney_1941}
B.~Bleaney and R.~A. Hull, Proc. R. Soc. London, Ser. A {\bf 178}(972), 86-92 (1941).

\bibitem{Yu_2010}
B.~Yu, M.~Liu, P.~W. Egolf, and A.~Kitanovski, Int. J. Refrig. {\bf 33}(6), 1029-1060 (2010).

\bibitem{Patterson_1985}
B.~R. Patterson and J.~A. Griffin, Mod. dev. powder metall. {\bf 15}(4), 279-288 (1985).

\bibitem{Dullien_1992}
F.A.L. Dullien, Porous Media. Fluid Transport and Pore Structure, Academic Press Inc., San Diego, USA. (1992).

\bibitem{Bahl_2012}
C.R.H. Bahl, D. Velazquez, K.K. Nielsen, K.	Engelbrecht, K.B. Andersen, R. Bulatova and N. Pryds, Appl. Phys. Lett. {\bf 100}(12), 121905 (2012).

\bibitem{comsol}
AB~Comsol~Multiphysics, Tegnergatan 23, SE-111 40 Stockholm, Sweden (2008).

\bibitem{Schenk_2001}
O.~Schenk, K.~G\"{a}rtner, W.~Fichtner, and A.~Stricker, Future Gen. Comp. Sys. {\bf 18}, 69-78 (2001).

\bibitem{Schenk_2002}
O.~Schenk and K.~G\"{a}rtner, Parallel Comp. {\bf 28}, 187-197 (2002).

\bibitem{Bjoerk_2010d}
R.~Bj\o{}rk, C.~R.~H. Bahl, and M.~Katter, J. Magn. Magn. Mater. {\bf 322}, 3882-3888, (2010).

\end{thebibliography}

\end{document}